\documentclass[aps,reprint,superscriptaddress,showpacs,showkeys,nofootinbib]{revtex4-1}

\usepackage{amsmath}
\usepackage{amssymb}
\usepackage{graphicx}

\begin{document}

\title{Impact of Coherent Neutrino Nucleus Scattering on\\Direct Dark Matter Searches based on CaWO$_4$ Crystals}

\newcommand{\mpi}{\affiliation{Max-Planck-Institut f\"ur Physik, D-80805 M\"unchen, Germany}}
\newcommand{\coimbra}{\affiliation{Departamento de Fisica, Universidade de Coimbra, P3004 516 Coimbra, Portugal}}
\newcommand{\vienna}{\affiliation{Institut f\"ur Hochenergiephysik der \"Osterreichischen Akademie der Wissenschaften, A-1050 Wien, Austria \\ and Atominstitut, Vienna University of Technology, A-1020 Wien, Austria}}
\newcommand{\tum}{\affiliation{Physik-Department, Technische Universit\"at M\"unchen, D-85748 Garching, Germany}}
\newcommand{\tuebingen}{\affiliation{Eberhard-Karls-Universit\"at T\"ubingen, D-72076 T\"ubingen, Germany}} 
\newcommand{\oxford}{\affiliation{Department of Physics, University of Oxford, Oxford OX1 3RH, United Kingdom}}
\newcommand{\wmi}{\affiliation{Walther-Mei\ss ner-Institut f\"ur Tieftemperaturforschung, D-85748 Garching, Germany}}
\newcommand{\lngs}{\affiliation{INFN, Laboratori Nazionali del Gran Sasso, I-67010 Assergi, Italy}}

\author{A.~G\"utlein}
\email{corresponding author, achim.guetlein@oeaw.ac.at}
\vienna

\author{G.~Angloher}
\mpi

\author{A.~Bento}
\coimbra 

\author{C.~Bucci}
\lngs

\author{L.~Canonica}
\lngs 

\author{A.~Erb}
  \tum
  \wmi

\author{F.~v.~Feilitzsch}
\tum 

\author{N.~Ferreiro~Iachellini}
\mpi

\author{P.~Gorla}
\lngs 

\author{D.~Hauff}
\mpi 

\author{J.~Jochum}
\tuebingen 

\author{M.~Kiefer}
\mpi

\author{H.~Kluck}
\vienna

\author{H.~Kraus}
  \oxford

\author{J.-C.~Lanfranchi}
\tum

\author{J.~Loebell}
\tuebingen

\author{A.~M\"unster}
\tum

\author{F.~Petricca}
\mpi 

\author{W.~Potzel}
\tum 

\author{F.~Pr\"obst}
\mpi

\author{F.~Reindl}
\mpi

\author{S.~Roth}
\tum 

\author{K.~Rottler}
\tuebingen 

\author{C.~Sailer}
\tuebingen 

\author{K.~Sch\"affner}
\lngs 

\author{J.~Schieck}
\vienna 

\author{S.~Sch\"onert}
\tum 

\author{W.~Seidel}
\mpi 

\author{M.~v.~Sivers}
\tum 

\author{L.~Stodolsky}
\mpi 

\author{C.~Strandhagen}
\tuebingen

\author{R.~Strauss}
\mpi 

\author{A.~Tanzke}
\mpi 

\author{M.~Uffinger}
\tuebingen 

\author{A.~Ulrich}
\tum 

\author{I.~Usherov}
\tuebingen 

\author{S.~Wawoczny}
\tum 

\author{M.~Willers}
\tum 

\author{M.~W\"ustrich}
\mpi 

\author{A.~Z\"oller}
\tum



\begin{abstract}
Atmospheric and solar neutrinos scattering coherently off target nuclei will be an important background source for the next generation of direct dark matter searches. In this work we focus on calcium tungstate as target material. For comparison with existing works we calculate the neutrino floor indicating which sensitivities can be reached before the neutrino background appears. In addition, we investigate the sensitivities of future direct dark matter searches using CRESST-II like detectors. Extending previous works we take into account achievable energy resolutions and thresholds as well as beta and gamma backgrounds for this investigation. We show that an exploration of WIMP-nucleon cross sections below the neutrino floor is possible for exposures of $\gtrsim 10$\,kg-years. In the third part we show that a first detection of coherent neutrino nucleus scattering of atmospheric and solar neutrinos using the same detectors and the backgrounds is feasible for exposures of $\gtrsim 50$\,kg-years.

\end{abstract}

\maketitle

\section{Introduction}\label{sec:Introduction}

The dynamics of galaxies and galaxy clusters \cite{DM1, DM2, DM3} give strong hints for the existence of dark matter. In addition, the precise measurements of the temperature fluctuations of the cosmic microwave background are well described by a contribution of $\sim 27$\,\% \cite{Planck} of cold dark matter to the overall energy density of the universe. However, the nature of dark matter remains unclear. Several direct dark matter searches \cite{CresstResults, SuperCDMS, EDELWEISS, LUX, XENON, DAMA, CoGeNT} aim at a detection of Weakly Interacting Massive Particles (WIMPs) \cite{DM1, DM2} scattering off nuclei.

In this work we focus on low-temperature detectors based on scintillating calcium tungstate (CaWO$_4$) crystals as those currently operated in the CRESST-II direct dark matter search \cite{TUM40Performance, CresstDetectors1, CresstDetectors2}. WIMPs are expected to scatter mainly off nuclei while the majority of the background interacts with electrons. The active suppression of these backgrounds is based on different amounts of scintillation light for electron- and nuclear-recoil events. However, atmospheric and solar neutrinos scattering off nuclei in the CaWO$_4$ crystal can mimic WIMP scatterings leading to an additional background source for future direct dark matter searches based on this technique.

In this work we discuss the impact of atmospheric and solar neutrinos on direct dark matter searches based on CaWO$_4$ as target material. First we focus on idealized detectors to calculate the neutrino floor indicating the best sensitivity which can be reached before the neutrino background appears. In the rest of the paper we focus on improved CRESST-II like detectors with achievable energy resolutions and thresholds as well as realistic $\beta$ and $\gamma$ backgrounds.

\section{Neutrino floor for CaWO$_4$}\label{sec:NeutrinoBg}
Coherent neutrino nucleus scattering (CNNS) \cite{CNNS} is a neutral current process\footnote{Neutrino-electron scattering as well as charged current interactions of the neutrino with the nucleus produce a charged lepton in the final state. Due to the suppression of electron-like events by the phonon-light technique, only CNNS is of importance as a possible background source for direct dark matter searches using CRESST-II like detectors.} of the weak interaction where a neutrino scatters elastically off a target nucleus via the exchange of a virtual Z$^0$ boson. For small transferred momenta the wavelength of the Z$^0$ is larger than the diameter of the nucleus. Thus, the neutrino scatters coherently off all nucleons.

Neutrinos scattering off nuclei in the detector can mimic WIMP scatterings leading to an additional background source for WIMP searches.
\begin{figure}[htb]
	\centering
	\includegraphics[width=0.49\textwidth]{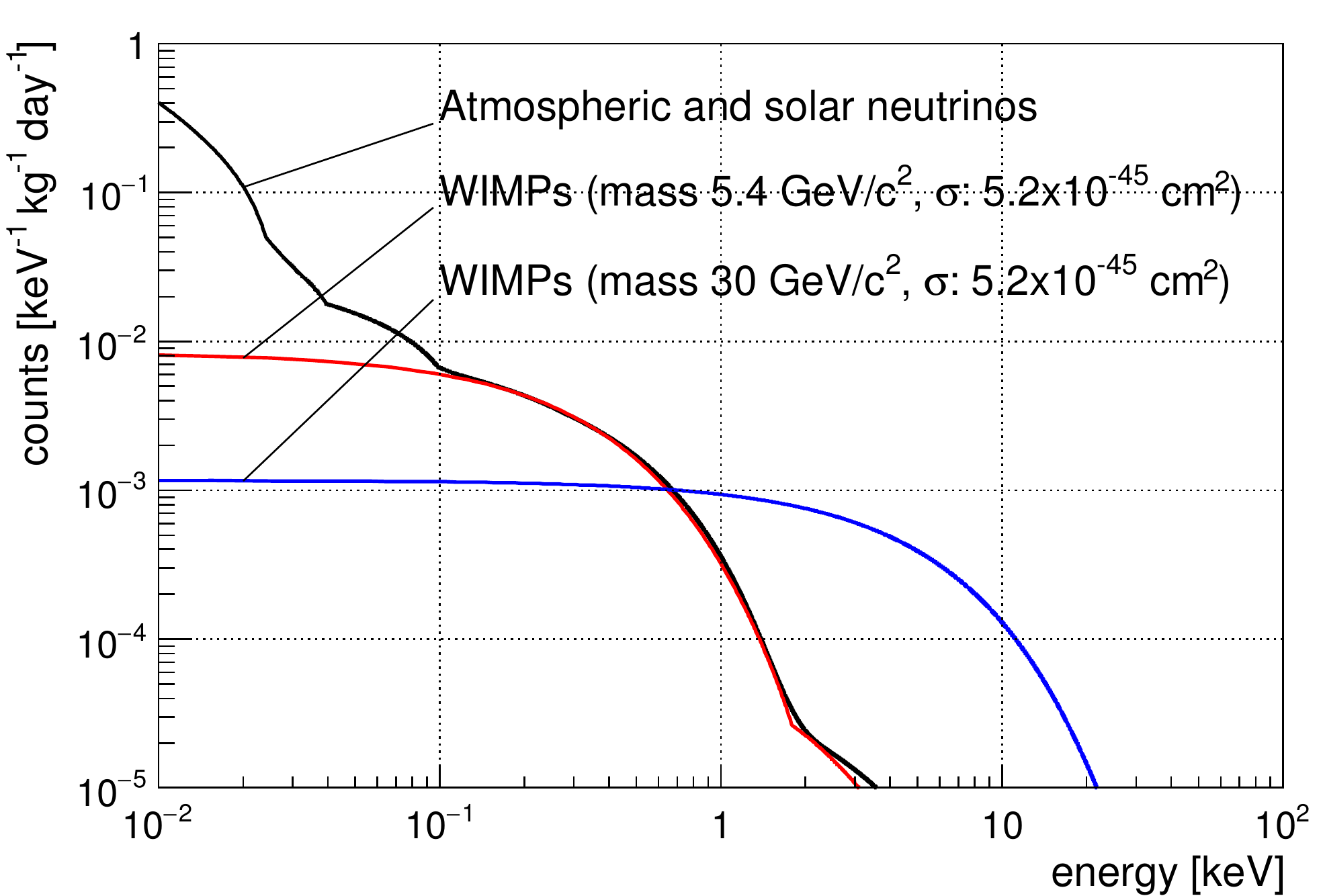}
	\caption{(Color online) Expected recoil-energy spectra for atmospheric and solar neutrinos as well as WIMPs with masses of 5.4\,GeV/c${}^2$ and 30\,GeV/c${}^2$ for CaWO$_4$ as target material. For a WIMP mass of 5.4\,GeV/c${}^2$ and a WIMP-nucleon scattering cross section of $5.2\cdot 10^{-45}$\,cm$^2$ the spectra of WIMPs and neutrinos are similar. Thus, for this WIMP scenario it is almost impossible to distinguish between neutrinos and WIMPs even for large exposures.}
	\label{fig:NuWimpComparison}
\end{figure}
FIG. \ref{fig:NuWimpComparison} shows the expected recoil-energy spectra for atmospheric and solar neutrinos as well as WIMPs with masses of 5.4\,GeV/c${}^2$ and 30\,GeV/c${}^2$ for CaWO$_4$ as target material. For a WIMP mass of $\sim 5.4$\,GeV/c${}^2$ and a WIMP-nucleon cross section of $\sim 5.2\cdot10^{-45}$\,cm$^2$ the expected spectra for neutrinos and WIMPs are very similar. Thus, for these WIMP masses it is nearly impossible to distinguish a WIMP signal from the neutrino background by their spectral shape and the sensitivity on a WIMP signal relies only on the knowledge of the neutrino rate.

For larger WIMP masses of $\sim 30$\,GeV/c${}^2$ the shapes of the neutrino and WIMP spectra are different. Thus, the sensitivity on a potential WIMP signal can be improved by taking the spectral shapes of signal and background into account. Such an analysis is only possible if enough neutrino events are observed to determine the spectral shape at energies $\gtrsim1$\,keV. For smaller exposures where only a few neutrino-background events are expected atmospheric and solar neutrinos scattering coherently off the target nuclei are a serious background source for all WIMP masses.

This neutrino background and its limitation for the sensitivities of direct dark matter searches have been studied in great detail during the last years \cite{NeutrinoBgLimit, NeutrinoBgMonroe, NeutrinoBgStrigari, NeutrinoBgGuetlein, GuetleinPhD, GuetleinAthen}.

Following the calculations in \cite{NeutrinoBgLimit} we estimate the neutrino floor for experiments based on CaWO$_4$. The neutrino floor is an optimistic estimation of the sensitivity on the WIMP-nucleon cross section which can be achieved before the neutrino background appears. The estimation is optimistic in the sense that idealized detectors are assumed, i.e., all detector specific properties like energy threshold and resolution, detection efficiency, and also limitations on the knowledge of the energy scale at low energies are neglected.


The thresholds used in this section are rather artificial analysis thresholds than actual limitations of a potential experiment. In addition to detector specific properties, all other backgrounds apart from neutrinos scattering coherently off target nuclei are neglected.

For an easier understanding of the following figures depicting the neutrino floor for CaWO$_4$ we are going to describe the method of \cite{NeutrinoBgLimit} to calculate the neutrino floor in the following.

For a given (analysis) threshold and exposure the expected number of neutrino events above this threshold can be calculated by integrating the expected recoil-energy spectra of atmospheric and solar neutrinos\footnote{In \cite{NeutrinoBgLimit} also the diffuse supernova neutrino background (DSNB) is included into the calculation. Since the DSNB is only a minor contribution and also the uncertainties on the flux and the energy spectrum are rather large, we decided not to include the DSNB into our calculations.}. For the calculation of the neutrino floor the exposure is adjusted in such a way that the expected number of neutrino events is one for a given threshold. Since the number of observed events follows a Poisson distribution, there is a probability of $\sim 37$\,\% to observe no event for an expected number of one neutrino event. This choise for the exposure leads also to an optimistic estimation of the sensitivity.

If an experiment observes neither background nor signal events, the exclusion limit, i.e., the upper limit on the WIMP-nucleon scattering cross section as a function of the WIMP mass, can be calculated using the Poisson distribution\footnote{Other methods based on (un)binned likelihoods will give the same result.}. The probability to observe no events is $10$\,\% for an expected number of $\sim 2.3$ WIMP events. Thus, a larger number of expected WIMP events is excluded with a probability of $90$\,\%. This upper limit on the number of expected WIMP events can be translated to an upper limit on the WIMP-nucleon scattering cross section as a function of the WIMP mass.

To calculate the neutrino floor the exclusion limits for different (analysis) thresholds are calculated under the assumption that neither signal nor background events are observed. Since the neutrino floor should be an optimistic estimate of the reachable sensitivity, for each WIMP mass it is given by the minimum of the upper limits of all thresholds.
\begin{figure}[htb]
	\centering
	\includegraphics[width=0.49\textwidth]{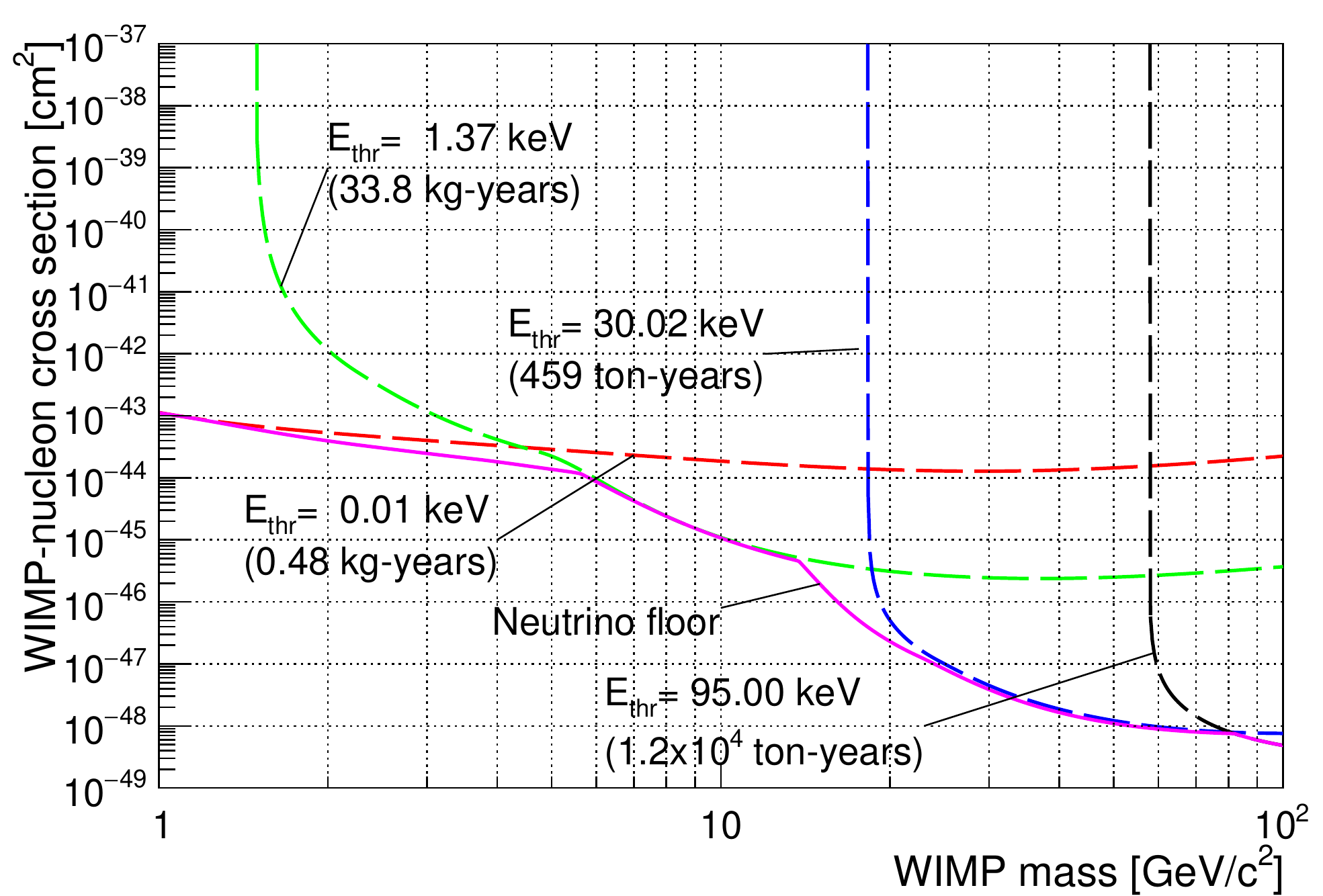}
	\caption{(Color online) The neutrino floor for CaWO$_4$ is shown as a solid (magenta) line. Exclusion limits for different thresholds and their corresponding exposures are shown as dashed lines. The kinks in the neutrino floor originate in a combination of the energy spectra of different neutrino sources and the recoil energies of the different nuclei of CaWO$_4$.}
	\label{fig:ExclusionLimits}
\end{figure}
FIG. \ref{fig:ExclusionLimits} shows exclusion limits for different thresholds and exposures as dashed lines. In addition, the neutrino floor is depicted as a magenta solid line. The kinks in the neutrino floor originate in a combination of the energy spectra of different neutrino sources and the recoil energies of the different nuclei of CaWO$_4$. The neutrino floor for high WIMP masses $\gtrsim 20$\,GeV/c${}^2$ are dominated by atmospheric neutrinos. For light WIMP masses the neutrino floor is dominated by solar neutrinos.

\section{Modeling CRESST-II like detectors}\label{sec:DetectorModel}

The detector modules currently operated in CRESST-II Phase 2 \cite{CresstResults, TUM40Performance} offer an active suppression of $\beta$ and $\gamma$ backgrounds leading to a background-free operation at energies $\gtrsim10$\,keV. In addition, the detectors have good energy resolutions ($\sim 0.1$\,keV \cite{TUM40Performance}) and low thresholds ($\sim 0.6$\,keV \cite{CresstResults, TUM40Performance}) leading to one of the best exclusion limits for WIMP masses below $\sim 3$\,GeV/c${}^2$ \cite{CresstResults}. In this section we describe models for the active background suppression, energy resolution and threshold, as well as $\beta$ and $\gamma$ backgrounds used for the calculations and simulations in sections \ref{sec:WimpSensitivity} and \ref{sec:NeutrinoDetection} (see \cite{TUM40Performance, TUM40Background} for further details).

\subsection{Working principle and active background suppression}

If a particle deposits energy in a CaWO$_4$ crystal, most of the deposited energy is converted into phonons, the rest into scintillation light escaping the CaWO$_4$ crystal\footnote{Depending on geometry and quality of the CaWO$_4$ crystal only a fraction of the scintillation light escapes the crystal while the rest is reabsorbed in the crystal adding up to the phonon signal.}. The phonon signal is measured by a transition edge sensor (TES) \cite{TUM40Performance, CresstDetectors1, CresstDetectors2} attached to the CaWO$_4$ crystal. The CaWO$_4$ crystal with the attached TES is called the phonon detector.

The scintillation light is detected with a separate light detector. The light detector consists of a silicon or silicon-on-sapphire plate to absorb the scintillation light \cite{TUM40Performance, CresstDetectors1, CresstDetectors2} and a TES to measure the scintillation-light signal, i.e., the phonon signal proportional to the absorbed scintillation light. Both, the phonon and light detectors are surrounded by a reflective foil to increase the amount of absorbed scintillation light \cite{TUM40Performance, CresstDetectors1, CresstDetectors2}.

The light yield, i.e., the ratio between scintillation light and phonon signal depends on the interaction type of the incident particle. The light yield is higher for electron recoils than for nuclear recoils. Thus, a simultaneous measurement of both signals allows for a discrimination between electron- and nuclear-recoil events on an event-by-event basis.

This phonon-light technique allows an active suppression of the majority of the $\beta$ and $\gamma$ background events, since WIMPs are expected to generate mainly nuclear recoils.

\subsection{Light-yield parametrization}\label{sec:LightYieldParametrization}

\begin{figure}[htb]
	\centering
	\includegraphics[width=0.49\textwidth]{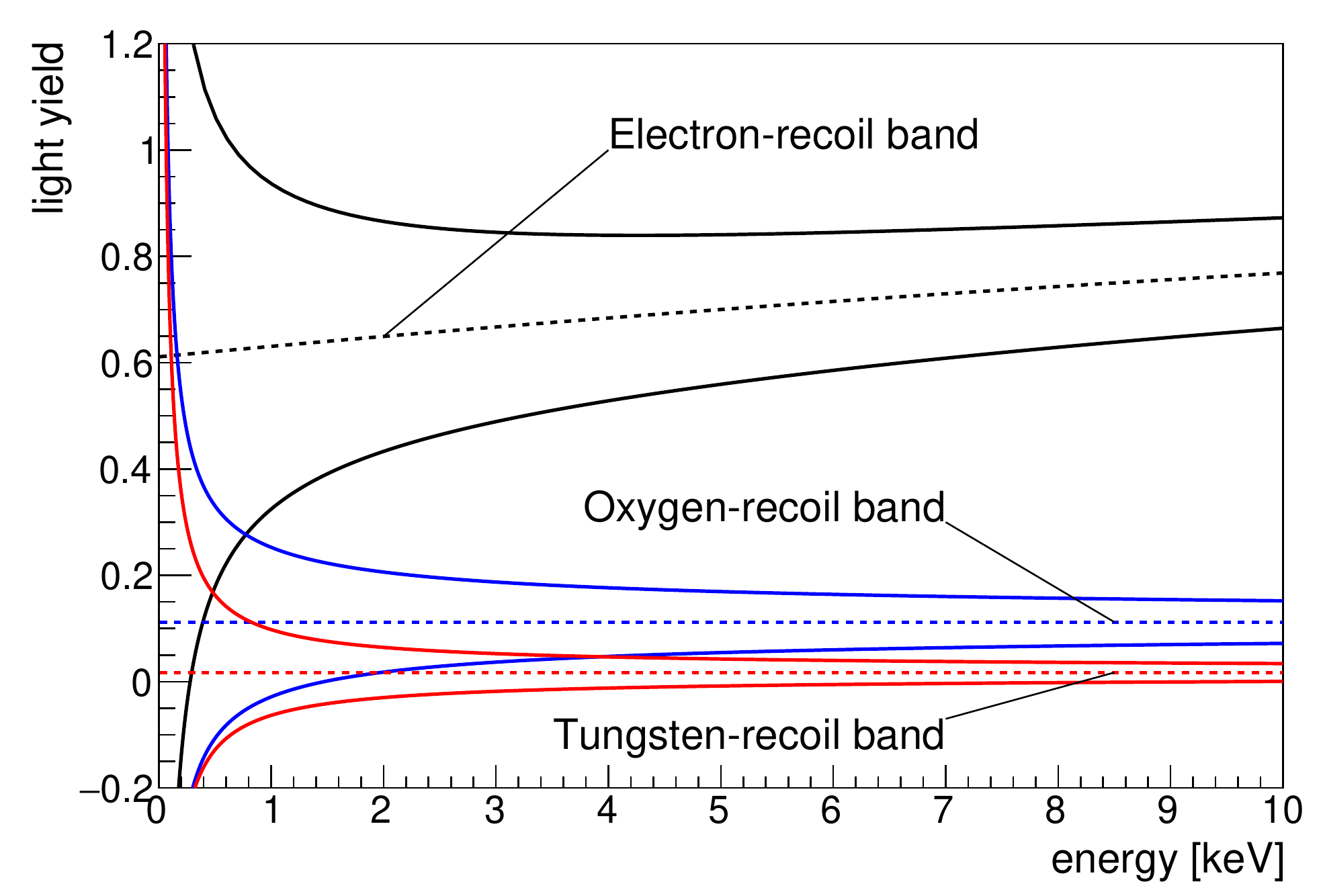}
	\caption{(Color online) The electron- (black), oxygen- (blue), and tungsten-recoil bands (red) are shown on the light yield versus energy plane. The calcium-recoil band lies between the oxygen- and tungsten-recoil bands. The means of the light-yield distributions are depicted as dashed lines. The solid lines mark the central 80\,\% intervals of the respective bands. The parameters used to calculate these band are given in the third column of TAB. \ref{tab:LightYieldPars}.}
	\label{fig:LightYield}
\end{figure}

The phonon-light technique to reject $\beta$ and $\gamma$ backgrounds is based on different amounts of scintillation light generated by different interactions. The measured data in the two channels (phonon and scintillation light) of CRESST-II like detector modules are typically displayed in a figure similar to FIG. \ref{fig:LightYield} where the abscissa is the deposited energy $E$ in the CaWO$_4$ crystal corrected for the energy of the scintillation light escaping the crystal \cite{CresstResults, TUM40Performance}. The ordinate is the light yield $LY$ which is defined as the ratio between phonon and scintillation-light signals \cite{CresstResults, TUM40Performance}. In order to compare the data form different detector modules the light yield is normalized to one for $122$\,keV $\gamma$s from a ${}^{57}$Co calibration source \cite{CresstResults, TUM40Performance}.

The light yield depends on the type of particle interaction and on the deposited energy. The mean light yield $\left<LY_{e, \gamma}(E)\right>$ for electrons and $\gamma$s\footnote{Small differences in the light yield of electrons and $\gamma$s have been found \cite{NonProportionality1, SabinPhD}. Since no well motivated model exists to describe this difference we approximately used the same light yield for $\gamma$s as for electrons.} and $\left<LY_{x}(E)\right>$ for $\alpha$-particles and nuclear recoils can be parametrized by \cite{TUM40Performance}:
\begin{eqnarray}
	\left<LY_{e, \gamma}(E)\right> & = & (p_0 + p_1E) \cdot (1 - p_2e^{-\frac{E}{p_3}})\label{equ:LightYieldElectron} \\
	\left<LY_{x}(E)\right> & = & (p_0 + p_1E) \cdot QF_x
\end{eqnarray}
where $p_0$ and $p_1$ account for the energy calibration of the detector modules. The second factor in equation (\ref{equ:LightYieldElectron}) is an empirical description of the fact that electrons with small energies generate less scintillation light due to an increased local energy loss \cite{NonProportionality1, NonProportionality2, SabinPhD}.
The quenching factors $QF_x$ for $\alpha$-particles and oxygen, calcium, and tungsten recoils are in principle energy dependent \cite{SabinPhD, RaimundQfPaper}. However, for energies $\lesssim40$\,keV considered in this work constant quenching factors are a good approximation \cite{TUM40Performance, RaimundQfPaper}.

Recoil bands are regions where events with the same physical origin (e.g. electron recoils or nuclear recoils) are located on the light yield versus energy plane. The light-yield distribution for a fixed energy is well described by a normal distribution \cite{RaimundQfPaper}. The finite width of a recoil band of given type $x = (e, \gamma, \alpha$, O, Ca, W) can be modelled by \cite{TUM40Performance, QfAlpha}:
\begin{equation}
	\sigma_x = \frac{1}{E} \sqrt{\sigma_L^2 + (\left<LY_x(E)\right>\sigma_P)^2 + S_1 E_L + S_2 E_L^2}\label{equ:LightYieldWidth}
\end{equation}
where $\sigma_L$ and $\sigma_P$ are the baseline fluctuations of light and phonon detector, respectively, and $E_L = \left<LY_x(E)\right>\cdot E$ is the energy of the scintillation-light signal. The term linear in $E_L$ with parameter $S_1$ accounts for photon statistics. The term quadratic in $E_L$ with parameter $S_2$ accounts for position dependencies in the light detector, negligible for energies $\lesssim 40$\,keV considered in this paper.

FIG. \ref{fig:LightYield} shows the light yield versus energy plane. The electron-, oxygen-, and tungsten-recoil bands are shown as black, blue and red lines, respectively. The calcium-recoil band lies between the oxygen- and tungsten-recoil bands. The means of all bands are shown as dashed lines. The solid lines mark the central 80\,\% intervals, i.e., the regions where 80\% of all electron- and oxygen-recoil events are expected. The parameters used to calculate these bands are given in the third column of TAB. \ref{tab:LightYieldPars}.

\subsection{Improved performance}\label{sec:ImprovedPerformance}

The current exclusion limit of CRESST-II Phase 2 \cite{CresstResults} was obtained with one detector module which is called TUM40 \cite{TUM40Performance}.
\begin{table}[htb]
	\centering
	\begin{tabular}{|l|r|r|}
		\hline
		\textbf{Parameter} & \textbf{TUM40} & \textbf{Improved detector}\\
		\hline
		$p_0$ & 0.938 & 1 \\
		\hline
		$p_1$ [keV$^{-1}$] & $4.6\cdot 10^{-5}$ & 0 \\
		\hline
		\hline
		$p_2$ & 0.389 & 0.389 \\
		\hline
		$p_3$ [keV] & 19.34 & 19.34 \\
		\hline
		\hline
		$\sigma_{P}$ [keV] & 0.091 & 0.05 \\
		\hline
		$\sigma_{L}$ [keV] & 0.269 & 0.04 \\
		\hline
		$S_1$ [keV] & 0.256 & 0.085 \\
		\hline
	\end{tabular}
	\caption{Values for the parameters used in the light-yield parametrization for two detectors. The second column (TUM40) lists the values for the TUM40 detector module \cite{TUM40Performance, CresstResults} operated in CRESST-II Phase 2. In the last column the corresponding values for the improved detector module discussed in this work are listed.}
	\label{tab:LightYieldPars}
\end{table}
The parameters for the light-yield pa\-ra\-metri\-za\-tion discussed in the present work are given in TAB. \ref{tab:LightYieldPars} \cite{TUM40Performance}.

The following studies are based on a detector module with an improved performance\footnote{We believe that it is possible to reach this improved performance during the next couple of years.}. The corresponding values for the light-yield parametrization are given in the last column of TAB. \ref{tab:LightYieldPars}. To obtain these values we assumed that it is possible to decrease the widths $\sigma_{P}$ and $\sigma_{L}$ of the baseline fluctuations of the phonon and the light detector by a factor of two. In addition, we assumed that the amount of detected scintillation light can be increased by a factor of three leading to a further decrease of the width $\sigma_{L}$ and also of the parameter $S_1$ by a factor of three.

\subsection{Background model and energy resolution}\label{sec:Background}

The background model used for this work was taken from \cite{TUM40Background}. This semi-empirical model describes the $\beta$ and $\gamma$ backgrounds observed with the TUM40 module operated in CRESST-II Phase 2. The average rate in the energy region $\lesssim40$\,keV is $\sim 3$\,keV$^{-1}$\,kg$^{-1}$\,day$^{-1}$ \cite{TUM40Background}. For the following studies we assume that this background rate can be decreased by a factor of 100. This might be realized by re-crystallization of CaWO$_4$ crystals to reduce intrinsic contamination and additional active and passive shielding close to the detectors.

In principle, the energy resolution is energy-dependent. However, for the energy region $\lesssim40$\,keV discussed in this work a constant value is a good approximation. Thus, we use $\sigma_{P}$ as energy resolution for the complete energy region considered for this work.

\subsection{Detection efficiency}

We used a two-component model for the detection efficiency, i.e., the probability that an event is recorded and survives all data-quality cuts. We assumed an energy-independent efficiency of 80\,\% due to dead time (i.e., time intervals with unstable detector response), data-quality cuts, and coincidences with a muon-veto system or other detector modules. In addition, we used an error function to model the trigger efficiency of the phonon detector. In CRESST-II the light detector response is always recorded if the phonon detector triggers. Thus, the trigger efficiency of the light-detector can be neglected. The mean of the error function\footnote{i.e., the energy where the error fuction has a value of 0.5.} for the trigger efficiency can be interpreted as the energy threshold of the phonon detector. We used a value of $5\sigma_{P} = 0.25$\,keV for the energy threshold. This value is typically chosen to avoid triggers from electronic and microphonic noise. The width of the trigger efficiency is given by the energy resolution for very low energies. For our model we used $\sigma_{P} = 0.05$\,keV as energy resolution for the complete energy region $\lesssim40$\,keV.

\section{Sensitivity for WIMP-nucleon scatterings}\label{sec:WimpSensitivity}

\begin{figure}[htb]
	\centering
	\includegraphics[width=0.49\textwidth]{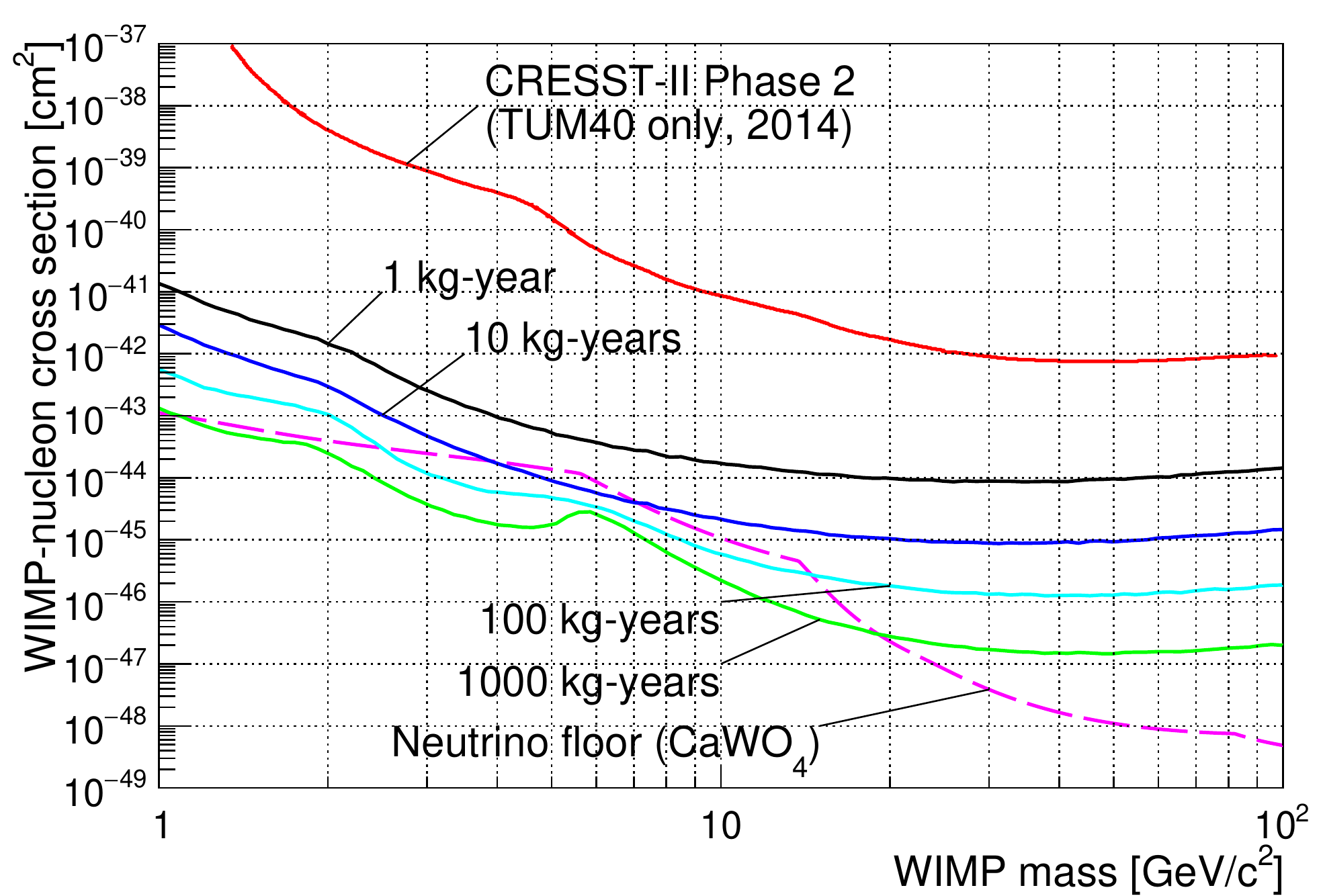}
	\caption{(Color online) The exclusion limits (90\,\% confidence level) of mock-data sets with exposures of 1 (black, solid), 10 (blue, solid), 100 (cyan, solid), and 1000\,kg-years (green, solid). The neutrino floor for CaWO$_4$ (see section \ref{sec:NeutrinoBg}) is shown as a (magenta) dashed line, the current limit by the CRESST-II experiment \cite{CresstResults} is shown as a red solid line.}
	\label{fig:SimExclusionLimits}
\end{figure}

In this section the sensitivity of the improved detector module on the cross section for elastic spin-independent WIMP-nucleon scattering is studied. Therefore we simulated mock-data sets with $\beta$ and $\gamma$ backgrounds as well as neutrinos for different exposures\footnote{One mock-data set for each exposure shown in FIG. \ref{fig:SimExclusionLimits}}. For these simulations we used the models given in section \ref{sec:DetectorModel} for $\beta$ and $\gamma$ backgrounds as well as the light-yield distribution. In addition, the atmospheric and solar neutrino signals was taken from \cite{NeutrinoBgGuetlein}.
FIG. \ref{fig:SimExclusionLimits} depicts the exclusion limits (90\,\% confidence level) obtained by unbinned likelihood fits\footnote{All fits in section \ref{sec:WimpSensitivity} and \ref{sec:NeutrinoDetection} were performed as two-dimensional unbinned likelihood fits in the light yield versus energy plane.} for exposures of 1 (black, solid), 10 (blue, solid), 100 (cyan, solid), and 1000\,kg-years (green, solid). In addition, the neutrino floor for CaWO$_4$ as calculated in section \ref{sec:NeutrinoBg} is shown as a (magenta) dashed line and the current limit by the CRESST-II experiment \cite{CresstResults} is shown as a red solid line.

The sensitivity, i.e., the upper limit on the cross section for the elastic spin-independent WIMP-nucleon scattering improves almost linearly with exposure between 1 and 10\,kg-years since the neutrino background is not important for these exposures. For higher exposures the neutrino background becomes more and more important limiting the sensitivity especially for WIMP masses of $\sim5$\,GeV/c${}^2$. For this WIMP mass the energy spectra of WIMPs and (solar) neutrinos are similar (see section \ref{sec:NeutrinoBg} and FIG. \ref{fig:NuWimpComparison}) leading to a saturation of the achieved sensitivity. This saturation is below the simplified estimation of the neutrino floor as calculated in \cite{NeutrinoBgLimit} and section \ref{sec:NeutrinoBg}.

A comparison of the neutrino floor (dashed magenta line) and the exclusion limits obtained by likelihood fits shows that the neutrino floor is an simplified estimation of the sensitivities which can be reached before backgrounds have to be taken into account. For WIMP masses $\lesssim 3$\,GeV/c${}^2$ the exclusion limits are dominated by low energies, where the phonon-light technique is ineffective. This can be seen in FIG. \ref{fig:LightYield} where the electron and oxygen recoil-bands overlap at energies $\lesssim 1$\,keV. Thus, for these low WIMP masses the sensitivity is rather limited by $\beta$ and $\gamma$ backgrounds than the neutrino background.

For masses $\gtrsim 15$\,GeV/c${}^2$ WIMPs generate larger recoil energies $\gtrsim10$\,keV (see FIG. \ref{fig:NuWimpComparison}). At these energies the spectra of WIMPs and solar neutrinos are well separated leading to a suppression of the solar neutrino background. Thus, for higher WIMP masses the sensitivity on the WIMP-nucleon cross section is limited by exposure. The atmospheric neutrino background will limit the sensitivity for exposures $\gtrsim 10$\,tonne-years which will be difficult to reach.

In the WIMP mass region between $\sim 3$ and $\sim 15$\,GeV/c${}^2$ the (solar) neutrino background is most relevant for a realistic experiment based on CaWO$_4$ as target material and the phonon-light technique provided by CRESST-II like detectors. For this region the neutrino background is dominated by ${}^8$B neutrinos\footnote{Solar neutrinos produced in the reaction ${}^8\text{B}\rightarrow {}^8\text{Be} + e^{+} + \nu_e$} \cite{NeutrinoBgLimit}.

\section{Detection of coherent neutrino nucleus scattering}\label{sec:NeutrinoDetection}

One way to detect a signal in the presence of a known background is to reject the background-only hypothesis. There are several frequentist (e.g., \cite{MaxLikelihoodRatio}) and bayesian me\-thods (e.g., \cite{BayesTheorem}) for hypothesis testing. For this work we used the maximum likelihood ratio test, a frequentist me\-thod (see e.g. \cite{MaxLikelihoodRatio} for details).

We simulated 1000 mock-data sets for each exposure given in the first column of TAB. \ref{tab:DetectionPotential}. We used the models given in section \ref{sec:DetectorModel} for the background spectra and the light-yield distribution. The atmospheric and solar neutrino spectra were taken from \cite{NeutrinoBgGuetlein}. For the maximum likelihood ratio test we fitted a background-only model as well as a model containing background and a neutrino signal to those mock-data sets. All fits were performed as two-dimensional unbinned likelihood fits in the light-yield versus energy plane.

In contrast to the fits obtained in section \ref{sec:WimpSensitivity}, we excluded a potential WIMP signal from the fits in this section. The reason for this is that the expected spectra for neutrinos and WIMPs are very similar for WIMP masses of $\sim5$\,GeV/c${}^2$ (see FIG. \ref{fig:NuWimpComparison}) and can only be distinguished by their rate. The WIMP contribution would always compete with the neutrino contribution making it impossible to detect a neutrino signal.  



\begin{figure}[htb]
	\centering
	\includegraphics[width=0.49\textwidth]{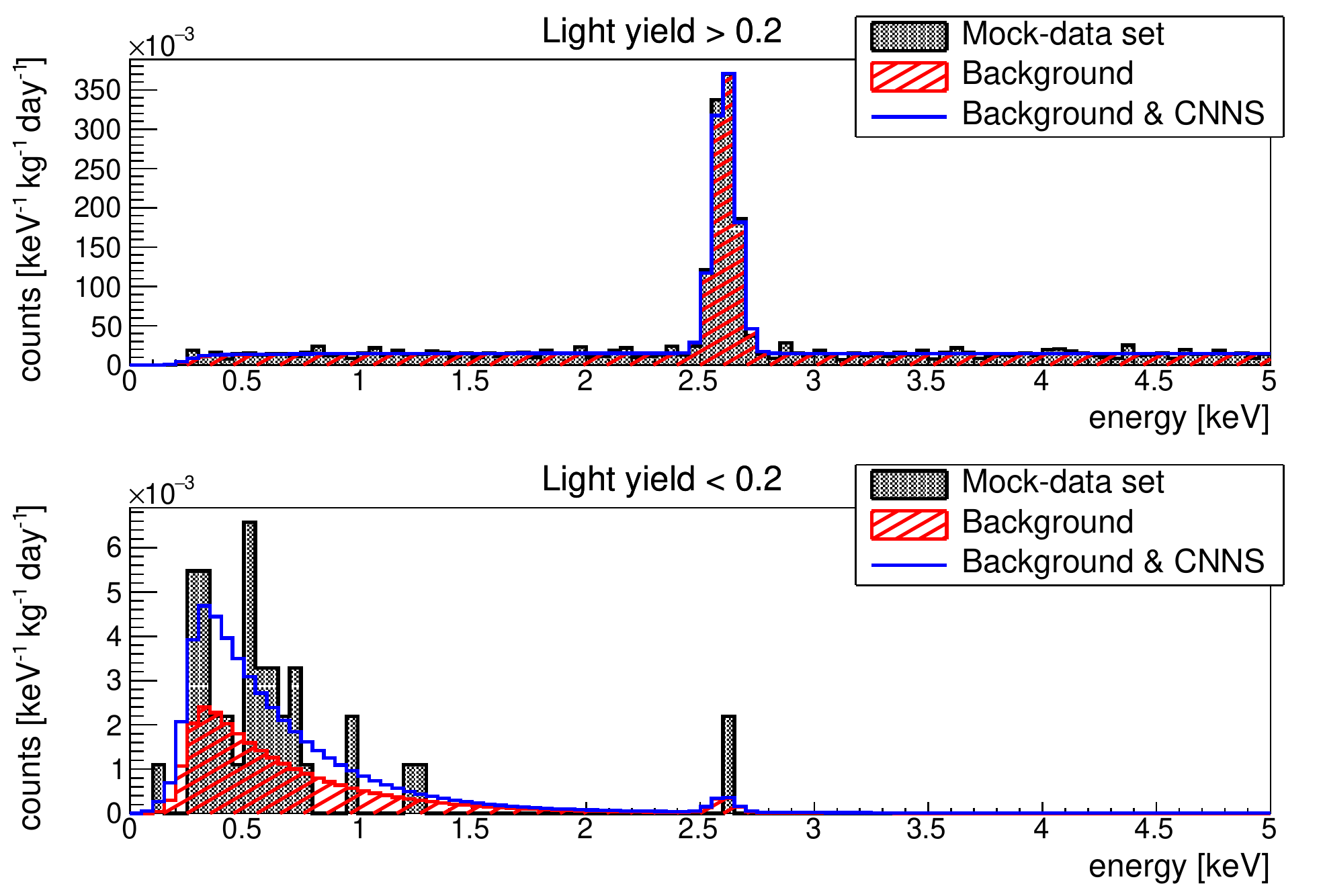}
	\caption{(Color online) One example of a mock-data set for an exposure of 50\,kg-years. The top panel shows a histogram (black, filled region) of all events with light yields larger than 0.2. The bottom panel shows a histogram with the remaining events (black, filled region). The colored histograms show the spectra of the background model (red, shaded region) and the sum of background and CNNS (blue, solid line). The peak at $\sim2.6$\,keV visible in both panels originates in the electron-capture decay of cosmogenic $^{179}$Ta.}
	\label{fig:FitBands}
\end{figure}
FIG. \ref{fig:FitBands} displays the result of a likelihood fit of the model with background and neutrinos to one mock-data set. For the simulation of this data set an exposure of 50\,kg-years was assumed. The top panel shows an energy histogram of all events with light yields larger than 0.2 and the bottom panel a histogram of the remaining events. In addition to the data (black, filled region) both panels also show spectra of the background model (red, shaded region) and the sum of electron/gamma backgrounds and CNNS (blue, solid line). The rates for both model histograms are obtained by a likelihood fit. The peak at $\sim2.6$\,keV visible in both panels originates in the electron-capture decay of cosmogenic $^{179}$Ta \cite{TUM40Background}.

As a result of the maximum likelihood ratio test the probability $P(\text{Data}|\text{Background-only})$ that the mock-data set shown in FIG. \ref{fig:FitBands} can be explained by a background-only model is $3.2\cdot 10^{-6}$. Thus, for this mock-data set a detection of CNNS at a confidence level of $99.9997$\,\% could be claimed.

In the following the detection potential is the probability to observe a data-set which allows a discovery of CNNS at the desired confidence level CL. To estimate this detection potential for different exposures we performed maximum likelihood ratio tests for all simulated mock-data sets.
\begin{table}[htb]
	\centering
	\begin{tabular}{|r||r|r|r|}
		\hline
		 & \multicolumn{3}{c|}{\textbf{Detection potential}} \\
		\hline
		\textbf{Exposure} & 99.9\,\% CL & 99.99\,\% CL & 99.9999\,\% CL \\
		\hline
		\hline
		10\,kg-years & 12.8\,\% & 7.6\,\% & 3.3\,\% \\
		\hline
		20\,kg-years & 28.9\,\% & 19.7\,\% & 9.3\,\% \\
		\hline
		30\,kg-years & 44.6\,\% & 29.9\,\% & 16.5\,\% \\
		\hline
		40\,kg-years & 61.1\,\% & 45.2\,\% & 23.7\,\% \\
		\hline
		50\,kg-years & 73.4\,\% & 57.9\,\% & 34.0\,\% \\
		\hline
		60\,kg-years & 80.8\,\% & 68.0\,\% & 42.1\,\% \\
		\hline
		70\,kg-years & 89\,\% & 79.4\,\% & 55.1\,\% \\
		\hline
		80\,kg-years & 91.7\,\% & 83.2\,\% & 64.9\,\% \\
		\hline
		90\,kg-years & 96.1\,\% & 90.6\,\% & 70.5\,\% \\
		\hline
		100\,kg-years & 97.4\,\% & 92.7\,\% & 77.2\,\% \\
		\hline
	\end{tabular}
	\caption{The detection potentials for different exposures. The detection potential is the probability to observe a data-set which allows a detection of CNNS at the desired confidence level CL. To estimate these detection potentials 1000 mock-data sets were simulated for each exposure and maximum likelihood ratio tests were performed for each data set.}
	\label{tab:DetectionPotential}
\end{table}
The estimated detection potentials for confidence levels CL of 99.9\,\%, 99.99\,\%, and 99.9999\,\% listed in TAB. \ref{tab:DetectionPotential} indicate that an exposure of at least 50\,kg-years is needed for a reasonable chance to detect CNNS on a confidence level of 99.99\,\% using the improved detector module described in section \ref{sec:ImprovedPerformance}.

\section{Conclusions}\label{sec:Conclusions}
Atmospheric and solar neutrinos scattering coherently off target nuclei can mimic WIMP-nucleus scatterings. Thus, for future direct dark matter searches coherent neutrino nucleus scattering (CNNS) can be expected to be an additional background source. In this work we focus on calcium tungstate (CaWO$_4$) as target material. For comparison with existing works, we calculated the neutrino floor which is an optimistic estimation of the sensitivities, which can be reached before the neutrino background appears. This optimistic estimation is achieved by assuming idealized detectors (infinite good energy resolution and threshold), arbitrary exposures and the absence of neutrino-background events although one is expected.

Extending existing works and for a more realistic study of the influence of the neutrino background we consider an improved CRESST-II like detector module with achievable energy resolution and threshold as well as realistic $\beta$ and $\gamma$ backgrounds. We show that it is possible to explore WIMP-nucleon cross sections below the neutrino floor by performing unbinned likelihood fits where the spectral shapes of WIMPs, neutrinos and $\beta$/$\gamma$ backgrounds are taken into account. However, for WIMP masses of $\sim5$\,GeV/c${}^2$ the solar ${}^{8}$B neutrino background leads to a saturation of the achievable sensitivity. For lighter WIMP masses the sensitivity is limited by $\beta$/$\gamma$ backgrounds and the power of the background suppression at low energies. For high WIMP masses $\gtrsim 20$\,GeV/c${}^2$ the sensitivity is limited by exposure. Atmospheric neutrinos will become an background source for this WIMP masses for exposures $\gtrsim10$\,ton-years.

In addition, we studied the potential for a first observation of CNNS using CRESST-II like detector modules with improved performance. We estimated the potential for a first detection of CNNS by performing maximum-likelihood ratio tests for simulated mock-data sets for different exposures. For an exposure of 50\,kg-years we obtained for $\sim 58$\,\% of the mock-data sets a discovery of CNNS at a confidence level of 99.99\,\%. Thus, a detection of CNNS will be in reach of a future experiment based on CaWO$_4$ as target material, if the assumed improvements on the detector performance and the background level can be achieved.

\section*{Acknowledgements}

This work was supported by funds of the German Federal Ministry of Science and Education (BMBF),
the DFG cluster of excellence 'Origin and Structure of the Universe', the Maier-Leibnitz-La\-bo\-ra\-to\-ri\-um (Garching), the Science and Technology Facilities Council (STFC) UK, as well as the Helmholtz Alliance for Astroparticle Physics.

\end{document}